\documentclass[a4paper, hyper, 11pt]{article}

\usepackage{jheppub} 

\usepackage{latexsym}
\usepackage{young}

\def\exp{{\rm exp}\,}
\def\d{\delta}

\def\a{\alpha}
\def\b{\beta}
\def\D{\Delta}
\def\L{\Lambda}
\def\l{\lambda}
\def\e{\epsilon}

\def\CF{{\cal F}}

\def\CN{{\cal N}}
\def\CO{{\cal O}}

\def\cint{\frac{1}{2\pi i}\oint}

\def\centeron#1#2{{\setbox0=\hbox{#1}\setbox1=\hbox{#2}\ifdim
   \wd1>\wd0\kern.48\wd1\kern-.48\wd0\fi
   \copy0\kern-.48\wd0\kern-.48\wd1\copy1\ifdim\wd0>\wd1
   \kern.48\wd0\kern-.48\wd1\fi}}

\newcommand{\beq}{\begin{equation}}
\newcommand{\eeq}{\end{equation}}
\newcommand{\bea}{\begin{eqnarray}}
\newcommand{\eea}{\end{eqnarray}}
\newcommand{\ba}{\begin{array}}
\newcommand{\ea}{\end{array}}

\newcommand{\p}{\partial}
\newcommand{\nn}{\nonumber}

\newcommand{\la}{\langle}
\newcommand{\ra}{\rangle}

\title{\boldmath Genus one correction to Seiberg-Witten prepotential from $\beta$-deformed matrix model}

 \author{Jong-Hyun Baek}
  \affiliation{Department of Physics, College of Science, Yonsei University, Seoul 120-749, Korea}

\emailAdd{jbaek@yonsei.ac.kr}

\abstract{We study $\beta$-deformed matrix models with Penner type potentials, which correspond to $\CN=2$ $SU(2)$ supersymmetric gauge theories with $N_F=2,3$, and $4$ flavors. We compute explicitly the genus one corrections to the free energy of the matrix model and show that they match the corresponding results obtained from the Nekrasov partition function.}

\begin{document}
\maketitle
\flushbottom

\section{Introduction}
%

The past several years have seen much progress in the study of $\CN=2$ superconformal field theories (SCFTs). It was shown that the compactification of M5-branes on a Riemann surface with punctures gives rise to a class of $\CN=2$ SCFTs \cite{Gaiotto:2009we}. Matter hypermultiplets of these SCFTs are related to the puctures on the Riemann surface. Elementary S-duality transformations are related  to different sewings of the same Riemann surface.

Moreover, it was proposed by Alday, Gaiotto and Tachikawa (AGT) \cite{Alday:2009aq} that the Nekrasov partition function of a class of  $\CN$=2 $SU(2)$ quiver SCFTs is identified with a chiral half of the correlation function of Liouville field theory on the corresponding Riemann surfaces. Specifically, the perturbative part of the partition function corresponds to the Liouville three-point function of primary operators and the instanton part of the partition function coincides with the conformal block generated by the Virasoro algebra.

In \cite{Wyllard:2009hg}, the AGT relation was generalized to $\CN=2$ SCFTs with gauge group $SU(N_c)$, relating them to the $ A_{N_c-1}$ conformal Toda theory. \cite{Mironov:2009by} Also, the instanton partition function of pure super Yang-Mills, which is the non-conformal limit of SCFTs obtained by sending the mass of the matter hypermultiplets to infinity, was identified with the norm of a coherent state, called Whittaker state, in Liouville field theory. \cite{Gaiotto:2009ma}, \cite{Marshakov:2009gn} Numerous works have been done investigating various aspects of this surprising relation.

In particular, Dijkgraaf and Vafa (DV) \cite{Dijkgraaf:2009pc} observed that the classical spectral curve of a Penner type matrix model can be interpreted as the Seiberg-Witten curve \cite{Seiberg:1994rs}, \cite{Seiberg:1994aj} of $\CN=2$ SCFT. They geometrically engineered the gauge theory with a suitable local Calabi-Yau geometry and employed the large $N$ duality of the B-model topological string to describe the gauge theory with the matrix model. Given the CFT description of the matrix model, where the number of insertion of screening operators in Toda theory is the rank $N$ of the matrices, \cite{Marshakov:1991gc}, \cite{Kharchev:1992iv}, it was suggested that the matrix model bridges between the gauge theory and the Liouville/Toda CFT. Matrix models corresponding to gauge theories with less number of flavors are also proposed in \cite{Eguchi:2009gf}. These matrix models in the context of the AGT relation have attracted much interest. \cite{Itoyama:2009sc}-\cite{Billo:2013fi}

According to the DV proposal, the coupling $g_s$ and the $\Omega$ background parameter $\e=\e_1+\e_2=g_s(\sqrt{\b}-1/\sqrt{\b})$ are of the same order in $g_s$. Thus, we double expand the free energy in $g_s$ and $\e$, fixing the ratio $Q=\e/g_s$ or equivalently the central charge of Liouville theory. \cite{Itoyama:2011mr} In other words, we consider the free energy as
\bea
F &=& \sum_{k,l\geq0}g_s^{2k} \e^l F_{k,l} \nn\\[2mm]
&=& F_{0,0} + \e F_{0,1} + \e^2 F_{0,2} + g_s^2F_{1,0} + \cdots \,.
\eea
In this paper, we will be computing the genus one part $F_{0,2}$ of the free energy in the $\b$-deformed matrix model. It has been shown in \cite{Eguchi:2010rf} that the genus zero free energy of the matrix model agrees with that of the $SU(2)$ gauge theory. This evidence was extended to the half genus case of the $\b$-deformed matrix model in \cite{Nishinaka:2011aa}. The genus one correction in the ordinary matrix model, i.e. $\b=1$, was studied in \cite{arXiv:0912.2988}.

In section \ref{mat}, we give a quick review of the $\b$-deformed matrix model with logarithmic potentials and derive the spectral curves from the loop equation. In the following three sections, we explicitly calculate the genus one part of the free energy of the matrix models for $N_F = 2,3,$ and $4$. In section \ref{sum}, we summarize and discuss the results. In Appendix \ref{Integral}, we present the detailed computation of integrals used in obtaining the free energy for $N_F=3$ and 4 cases. In appendix \ref{Nek}, the Nekrasov instanton partition function is reviewed and the free energy is calculated from them.

\section{$\b$-deformed Matrix Model and Spectral Curve}\label{mat}

\subsection{Penner type Matrix Model}
The partition function of $\b$-deformed matrix model is given by
\beq\label{part}
Z = \int\left[\prod_{I=1}^{N} d\l_I\right] \D^{2\b}\, \exp\left[\frac{\sqrt{\b}}{g_s}\sum_{I=1}^{N} V(\l_I)\right] \,,
\eeq
where $\D = \prod_{I<J}(\l_I-\l_J)$ is the Vandermonde determinant.
The $\Omega$ background parameters are related to the matrix model parameters through
\beq
\e_1 = g_s\sqrt{\b}, \qquad \e_2 = - \frac{g_s}{\sqrt{\b}}\,.
\eeq
The potentials for $N_F=2,3, 4$ cases are given by
\bea
V(z)_{N_F=2} &=& \left(2\mu_3 + \e\right) \log z + \L_2\left(z+\frac{1}{z}\right) \,,\label{po2}\\
V(z)_{N_F=3} &=& (2\mu_3 + \e)\log z + 2m_1 \log(z-1) -\frac{\L_3}{z} \,,\label{po3}\\[3mm]
V(z)_{N_F=4} &=& (2m_0 + \e)\log z + 2m_1\log(z-1) + 2m_2\log(z-q) \,,\label{po4}
\eea
where the mass parameters $m_0, m_1, m_2$ with additional $m_{\infty}$, are related to the four anti-fundamental hypermultiplet masses $\mu_i$ of the gauge theory by
\beq
\mu_1 = m_1+m_{\infty}, \quad \mu_2 = m_1-m_{\infty}, \quad \mu_3 = m_2+m_0, \quad \mu_4 = m_2-m_0 \,
\eeq
The $\L_2$ and $\L_3$ are dimensionful parameters and correspond to the dynamical scales of the gauge theory, whereas $q$ is dimensionless and is identified with the exponential of the UV coupling of the SCFT. The potential for $N_F=3$ case is derived from $N_F=4$ potential by taking $\mu_4\rightarrow\infty$ with fixed $\L_3 \equiv q\mu_4$. The potential for $N_F=2$ is obtained from the $N_F=3$ potential by sending $\mu_2\rightarrow\infty$ while $\L_2^2 \equiv \L_3 \mu_2$ is fixed. The neutrality condition in Liouville theory states that
\beq\label{nc}
\mu_1+\mu_3+g_s\sqrt{\b}N = 0 \,.
\eeq

The free energy of the matrix model is defined to be
\beq
F \equiv g_s^2 \log Z = -\e_1\e_2 \log Z \,,
\eeq
and it can be expanded in $g_s$ and $\e=\e_1+\e_2$,
\beq
F = \sum_{k,l\geq0}g_s^{2k} \e^l F_{k,l} \,.
\eeq
In ordinary matrix model, where $\b=1$, the parameter $\e$ vanishes so only those terms with $l=0$ survive. In this paper, we compute the leading $\e^2$-correction $F_{0,2}$ to the free energy.

\subsection{Spectral Curves from Loop Equation}

We define $n$-point connected resolvent by
\beq
W(z_1,\cdots,z_n) = \b\left(\frac{g_s}{\sqrt{\b}}\right)^{2-n}\left\la\sum_{I_1}\frac{1}{z_1-\l_{I_1}}\cdots\sum_{I_n}\frac{1}{z_n-\l_{I_n}}\right\ra_c \,.
\eeq
The loop equation, which is obtained from the variation of the partition function (\ref{part}) by $\d\l_I = \frac{\a}{\l_I-z}$ for small $\a$, is given by \cite{Nishinaka:2011aa}, \cite{Chekhov:2010xj},
\beq\label{loop}
g_s^2 W(z,z) + W(z)^2 + \e\,W'(z) + V'(z)W(z) - f(z) = 0 \,,
\eeq
where the function $f(z)$ is
\beq\label{f}
f(z) = g_s\sqrt{\b}\sum_{I}\left\la\frac{V'(z)-V'(\l_I)}{z-\l_I}\right\ra \,.
\eeq
We expand the resolvents and the potential as follows.
\beq
W(z_1,\cdots,z_n) = \sum_{k,l\geq0}g_s^k\e^l W_{k,l}(z_1,\cdots,z_n) \nn
\eeq
\beq
V(z) = V_0(z) + \e V_1(z) \nn
\eeq
Then, the loop equation (\ref{loop}) is expanded to
\bea\label{loopexp}
g_s^0\e^0: && W_{0,0}^2(z) + V'_0(z)W_{0,0}(z) - f(z) = 0 \nn\\[2mm]
g_s^0\e^1: && 2\,y_{0,0}(z)W_{0,1}(z) + W'_{0,0}(z) + V'_{1}(z)W_{0,0}(z) = 0 \\[2mm]
g_s^0\e^2: && 2\,y_{0,0}(z)W_{0,2}(z) + W_{0,1}^2(z) + W'_{0,1}(z) + V_1'(z)W_{0,1}(z)=0 \nn
\eea
up to $\e^2$, where we have defined
\beq\label{y0}
y_{0,0}(z) \equiv W_{0,0}(z) + \frac{1}{2}V'_0(z) = \frac{1}{2}\sqrt{V'_0(z)^2 + 4f(z)} \,,
\eeq
which is the leading order term in the double expansion of the spectral curve,
\beq
y(z) \equiv W(z) + \frac{1}{2}V'(z) = \sum_{k,l\geq0}g_s^{k}\e^l y_{k,l}(z)\,.
\eeq
It is easy to see from the expansion of the loop equation (\ref{loop}) that $W_{1,0}(z)$ and $W_{1,1}(z)$ vanish.

The filling fraction in the matrix model is identified with the Coulomb branch parameter $a$ of the gauge theory and is given by the integral of the one form $y(z)dz$ along the A-cycle of the spectral curve,
\bea
a &=& \frac{1}{2\pi i}\oint_A y(z)dz \,,\nn\\[2mm]
&=& \sum_{k,l\geq0} g_s^k\e^la_{k,l} \,,
\eea
where we have double expanded the vev.
To compute the period integral, we first express $y_{0,1}(z)$ and $y_{0,2}(z)$ in terms of $y_{0,0}(z)\equiv y_0(z)$ and the potential. From (\ref{loopexp}) and (\ref{y0}), we obtain
\bea
y_{0,1}(z) &=& W_{0,1}(z) + \frac{1}{2}V_1'(z) \nn\\[2mm]
&=& -\frac{y_0'}{y_0} + \frac{V_0'' + V_0'V_1'}{2y_0} \,,\label{y01}\\[3mm]
y_{0,2}(z) &=& W_{0,2}(z) \nn\\[2mm]
&=& - \frac{1}{2y_0}\bigg(W_{0,1}^2(z) + W'_{0,1}(z) + V_1'(z)W_{0,1}(z)\bigg) \nn\\[2mm]
&=& -\frac{1}{2y_0}\left(y_{0,1}^2 + y_{0,1}'- \frac{1}{4}V_1'^2 -\frac{1}{2}V_1''\right) \label{y02}\,.
\eea
Using the expression \eqref{y01} of $y_{0,1}(z)$, we write $y_{0,2}(z)$ as
\beq\label{y2}
y_{0,2}(z) = \frac{y_0'^2}{8y_0^3} - \frac{(V_0''+V_0'V_1')^2}{32y_0^3} + \frac{V_1'^2+2V_1''}{8y_0} + \frac{d}{dz}\left(\frac{2y_0'-V_0''-V_0'V_1'}{8y_0^2}\right)
\eeq
Thus, the subleading terms of the vector multiplet vev read
\bea
a_{0,1} &=& \frac{1}{2\pi i}\oint \frac{V_0'' + V_0'V_1'}{2y_0}dz \,,\label{a01}\\[3mm]
a_{0,2} &=& \frac{1}{2\pi i}\oint \left(\frac{y_0'^2}{8y_0^3} - \frac{(V_0''+V_0'V_1')^2}{32y_0^3} + \frac{V_1'^2+2V_1''}{8y_0} \right)dz \label{a02}\,,
\eea
where we have dropped from $a_{0,1}$ the shift caused by the total derivative of $\log y_0$. \cite{Nishinaka:2011aa}

\section{$N_F=2$ Model}\label{f2}

We first consider the matrix model which is dual to the $SU(2)$ gauge theory with $N_F=2$. The action is given by
\beq
V(z) = V_0(z) + \e V_1(z) = 2\mu_3\log z + \L\left(z+\frac{1}{z}\right) + \e\log z \,,
\eeq
where we have omitted the subscript from $\L_2$ in \eqref{po2}. The function $f(z)$ in \eqref{f} is evaluated to be
\beq
f(z) = \frac{c_1}{z}+\frac{c_2}{z^2} \,,
\eeq
with
\bea
c_1 &=& g_s\sqrt{\b}\left\la\sum_{I=1}^N\left(-\frac{2\mu_3+\e}{\l_I}+\frac{\L}{\l_I^2}\right)\right\ra = g_s\sqrt{\b}N\L = -(\mu_1+\mu_3)\L \,, \\[2mm]
c_2 &=& g_s\sqrt{\b}\L\left\la\sum_{I=1}^N\frac{1}{\l_I}\right\ra\,,
\eea
where the equation of motion $\la\sum_I V'(\l_I)\ra=0$ and (\ref{nc}) have been used in computing $c_1$.
The planar spectral curve (\ref{y0}) becomes
\beq\label{y0f2}
y_0(z)^2 = \frac{\L^2}{4}\frac{P_4(z)}{z^4} \,,
\eeq
where the quartic polynomial $P_4(z)$ is given by
\beq\label{p4f2p}
P_4(z) = z^4 - \frac{4\mu_1}{\L}z^3 + \frac{4}{\L^2}\left(\mu_3^2 + c_2 - \frac{\L^2}{2}\right)z^2 - \frac{4\mu_3}{\L}z +1 \,.
\eeq
From now on, we will take $\mu_1=\mu_3=m$ for simplicity. Then we have
\beq\label{p4f2}
P_4(z) = z^4 - \frac{4m}{\L}z^3 + \frac{4A}{\L^2}z^2 - \frac{4m}{\L}z +1
\eeq
with $A \equiv m^2 + c_2 - \L^2/2$. The leading order vector multiplet vev $a_{0,0} \equiv a_0$ and the half genus contribution $a_{0,1}$ in (\ref{a01}) are written as
\bea
a_0 &=& \frac{1}{2\pi i}\oint \frac{\L}{2}\frac{\sqrt{P_4(z)}}{z^2}dz \,, \label{a0f2}\\[3mm]
a_{0,1} &=& \frac{1}{2\pi i}\oint \frac{1}{2}\left(z+\frac{1}{z}\right)\frac{dz}{\sqrt{P_4(z)}} = -\frac{1}{2}\frac{\p a_0}{\p m} \label{a1f2}
\eea

We are going to express the genus one part $a_{0,2}$ of the vev in terms of the derivatives of $a_0$, which are given by
\bea
\frac{\p a_0}{\p A} &=& \frac{1}{2\pi i}\oint \frac{dz}{\L\sqrt{P_4}} \label{A2}\,,\\[2mm]
\frac{\p^2 a_0}{\p m^2} &=& \frac{1}{2\pi i}\oint \left(-\frac{2}{\L}\right)(z^2+1)^2\frac{dz}{P_4^{3/2}} \,,\label{mm2}\\[1mm]
\frac{\p^2 a_0}{\p m\p A} &=& \frac{1}{2\pi i}\oint \frac{2}{\L}(z^3+z)\frac{dz}{P_4^{3/2}} \,,\label{mA2}\\[1mm]
\frac{\p^2 a_0}{\p A^2} &=& \frac{1}{2\pi i}\oint \left(-\frac{2}{\L^3}\right)z^2\frac{dz}{P_4^{3/2}} \label{AA2}\,.
\eea
The $a_{0,2}$ in (\ref{a02}) consists of three terms and we will consider the three terms in turn. The first term is
\beq\label{ft2}
\frac{1}{2\pi i}\oint \frac{y_0'^2}{8y_0^3}dz = \frac{1}{2\pi i}\oint \frac{1}{\L}\left(\frac{z^2}{16}\frac{P_4'^2}{P_4^{5/2}} - \frac{z}{2}\frac{P_4'}{P_4^{3/2}} + \frac{1}{\sqrt{P_4}}\right)dz \,.
\eeq
The first integral on the right hand side can be written as
\bea
 \frac{1}{2\pi i}\oint \frac{z^2}{16\L}\frac{P_4'^2}{P_4^{5/2}}dz &=& \cint \frac{z^2}{16\L}\left(\frac{4}{3}\frac{d^2}{dz^2}\frac{1}{\sqrt{P_4}} + \frac{2}{3}\frac{P_4''}{P_4^{3/2}}\right)dz \nn\\[2mm]
 &=& \cint\frac{1}{6\L}\left(\frac{1}{\sqrt{P_4}} + \frac{z^2P_4''}{4P_4^{3/2}}\right)dz \label{ff2}\,,
\eea
where we have integrated by parts in the last equality. The explicit form of $P_4(z)$ in \eqref{p4f2} satisfies
\beq
z^2P_4'' = 3zP_4' + \frac{12m}{\L}(z^3+z) - \frac{16A}{\L^2}z^2 \,.
\eeq
Given the above relation and using the derivatives of $a_0$ in (\ref{A2}), (\ref{mA2}), (\ref{AA2}), we find (\ref{ff2}) to be
\beq
\frac{1}{2\pi i}\oint \frac{z^2}{16\L}\frac{P_4'^2}{P_4^{5/2}}dz = \frac{1}{6}\frac{\p a_0}{\p A} + \cint \frac{z}{8\L}\frac{P_4'}{P_4^{3/2}}dz + \frac{m}{4}\frac{\p^2a_0}{\p m\p A} + \frac{A}{3}\frac{\p^2a_0}{\p A^2} \,.
\eeq
Thus, we obtain the first term (\ref{ft2}) of $a_{0,2}$  as
\bea\label{fft2}
\frac{1}{2\pi i}\oint \frac{y_0'^2}{8y_0^3}dz &=& \frac{7}{6}\frac{\p a_0}{\p A} + \frac{m}{4}\frac{\p^2a_0}{\p m\p A} + \frac{A}{3}\frac{\p^2a_0}{\p A^2} + \cint \frac{-3z}{8\L}\frac{P_4'}{P_4^{3/2}}dz \nn\\[2mm]
&=& \frac{5}{12}\frac{\p a_0}{\p A} + \frac{m}{4}\frac{\p^2a_0}{\p m\p A} + \frac{A}{3}\frac{\p^2a_0}{\p A^2} \,,
\eea
where we have used in the last step
\beq
\cint z \frac{P_4'}{P_4^{3/2}}dz = \cint \frac{2}{\sqrt{P_4}}dz = 2\L\frac{\p a_0}{\p A} \,.
\eeq

The second and third terms of $a_{0,2}$ in (\ref{a02}) are given by
\bea
\cint \frac{(V_0''+V_0'V_1')^2}{32y_0^3}dz &=& \cint \frac{1}{4\L}(z^2+1)^2\frac{dz}{P_4^{3/2}}dz \nn\\[2mm]
&=& -\frac{1}{8}\frac{\p^2 a_0}{\p m^2} \,,\label{fst2}\\[2mm]
\cint \frac{V_1'^2+2V_1''}{8y_0}dz &=& \cint \left(-\frac{1}{4\L}\right)\frac{dz}{\sqrt{P_4}} \nn\\[2mm]
&=& -\frac{1}{4}\frac{\p a_0}{\p A} \label{ftt2}\,.
\eea
Plugging (\ref{fft2}), (\ref{fst2}), and (\ref{ftt2}) into (\ref{a02}), we find
\beq
a_{0,2} = \frac{1}{6}\frac{\p a_0}{\p A} + \frac{m}{4}\frac{\p^2a_0}{\p m\p A} + \frac{A}{3}\frac{\p^2a_0}{\p A^2} + \frac{1}{8}\frac{\p^2 a_0}{\p m^2} \,.\label{a02f2}
\eeq

The planar vev (\ref{a0f2}) can be computed by doing elliptic integral and expanding hypergeometric function, which was done in \cite{Eguchi:2010rf}.
\begin{eqnarray}
a_{0} &=& \sqrt{A}\left(1-\frac{m^2}{4A^2}\Lambda^2 - \frac{(A^2 - 6m^2A + 15m^4)}{64A^4}\Lambda^4 - \frac{5(3m^2 A^2 - 14m^4A + 21 m^6)}{256A^6}\Lambda^6\right.
\nonumber \\[2mm]
&& \qquad \left. - \frac{15(A^4 - 28m^2A^3 + 294 m^4 A^2 - 924m^6A + 1001m^8)}{16384A^8}\Lambda^8 + \mathcal{O}(\Lambda^{10})
\right)\,.
\label{fa0f2}
\end{eqnarray}
The half genus part (\ref{a1f2}) is given by \cite{Nishinaka:2011aa}
\begin{eqnarray}
a_{0,1} &=& \frac{m}{4 A^{3/2}}\Lambda^2-\frac{3m \left(A-5 m^2\right)}{32 A^{7/2}}\Lambda^4+\frac{5 \left(3 A^2 m-28 A m^3+63 m^5\right)}{256 A^{11/2}} \Lambda^6
\nonumber\\[3mm]
&& \quad -\frac{105 \left(m \left(A^3-21 A^2 m^2+99 A m^4-143 m^6\right)\right) }{4096 A^{15/2}}\Lambda^8 + \cdots\,.
\label{fa1f2}
\end{eqnarray}
The genus one part (\ref{a02f2}) is
\bea
a_{0,2} &=& -\frac{A+m^2}{16 A^{5/2}}\Lambda ^2+\frac{A^2-30 Am^2-35 m^4}{128 A^{9/2}}\Lambda ^4-\frac{5  \left(3 A^3-49 A^2 m^2+147 Am^4+231 m^6\right)}{1024 A^{13/2}}\Lambda ^6 \nn\\[3mm]
&& \quad +\frac{35  \left(A^4-126 A^3 m^2+792A^2 m^4-858 A m^6-2145 m^8\right)}{16384 A^{17/2}}\Lambda ^8 + \cdots \,.
\eea
We invert the equation $a=a_0 + \e a_{0,1} + \e^2a_{0,2}$ to find $A$ in terms of $a$:
\bea
A &=& \left(a^2 + \frac{m^2}{2 a^2}\Lambda ^2 +\frac{a^4-6a^2m^2+5m^4}{32a^6}\Lambda ^4 +\frac{5 a^4 m^2-14 a^2 m^4+9 m^6}{64a^{10}}\Lambda ^6 \right.\nn\\[3mm]
&&\qquad\qquad + \left.\frac{5 a^8-252a^6m^2+1638 a^4 m^4-2860 a^2 m^6+1469 m^8}{8192 a^{14}}\Lambda ^8 + \cdots\right) \nn\\[4mm]
&& +\e\left(-\frac{ m}{2 a^2}\Lambda ^2 + \frac{3 a^2 m-5 m^3}{16 a^6}\Lambda ^4 - \frac{5 a^4
   m-28 a^2 m^3+27 m^5}{64 a^{10}}\Lambda ^6 \right. \nn\\[3mm]
&&\qquad\qquad  +  \left.\frac{63 a^6 m-819 a^4 m^3+2145 a^2 m^5-1469
   m^7}{2048 a^{14}}\Lambda ^8 + \cdots \right) \nn\\[4mm]
&& +\e^2\left( \frac{a^2 + m^2}{8
   a^4}\Lambda ^2 - \frac{a^4-21 m^4}{64 a^8}\Lambda ^4  + \frac{5 a^6-14 a^4 m^2-123 a^2 m^4+220 m^6}{256 a^{12}}\Lambda ^6  \right.   \nn\\[3mm]
&&\qquad  +  \left.\frac{
   -21 a^8+738 a^6 m^2+3080 a^4 m^4-19266 a^2 m^6+18445 m^8}{8192 a^{16}}\Lambda ^8 + \cdots \right)
\eea

In order to obtain the free energy, we consider the relation
\beq
\L\frac{\p}{\p \L}F = g_s\sqrt{\b}\L\sum_{I}\left\la\l_I + \frac{1}{\l_I}\right\ra = g_s\sqrt{\b}\L\la\sum_I \l_I\ra + c_2 \,.
\eeq
The vev $\la\sum_I\l_I\ra$ can be evaluated by looking at large $z$ behavior of the planar resolvent $W_{0,0}(z)\approx g_s\sqrt{\b}N/z + g_s\sqrt{\b}\la\sum_I\l_I\ra/z^2 + \cdots$ in the loop equation \eqref{loopexp}.
\beq
g_s\sqrt{\b}\L\la\sum_I\l_I\ra = c_2 + (\mu_1^2-\mu_3^2) + \cdots
\eeq
For $\mu_1=\mu_3=m$, we have
\beq
\L\frac{\p}{\p \L}F = 2c_2  = 2(A - m^2) + \L^2
\eeq
up to $g_s^2$. By integrating in $\L$, we obtain the free energy $F = F_{0,0}+\e F_{0,1}+\e^2 F_{0,2}+\cdots$ for $N_F=2$ model. The planar part is given by
\begin{eqnarray}
 F_{0,0} &=& 2(a^2-m^2)\log \Lambda + \frac{a^2+m^2}{2a^2}\Lambda^2+ \frac{a^4-6a^2m^2 + 5m^4}{64a^6}\Lambda^4
\nonumber\\[3mm]
&& + \frac{5a^4m^2 - 14a^2m^4 +9m^6}{192a^{10}}\Lambda^6
\nonumber\\[3mm]
&&+ \frac{5 a^8-252 a^6 m^2+1638 a^4 m^4-2860 a^2 m^6+1469 m^8}{32768 a^{14}}\Lambda^8 + \cdots\,,
\end{eqnarray}
the half genus part by
\begin{eqnarray}
F_{0,1} &=&   - \frac{m}{2a^2}\Lambda^2 + \frac{3a^2m-5m^3}{32a^6}\Lambda^4 -\frac{5 a^4 m-28 a^2 m^3+27 m^5}{192 a^{10}}\Lambda^6
\nonumber\\[3mm]
&&\quad +\frac{63 a^6 m-819 a^4 m^3+2145 a^2 m^5-1469 m^7}{8192a^{14}}\Lambda^8 + \cdots\,,
\end{eqnarray}
and the genus one part by
\bea
F_{0,2} &=& \frac{a^2+m^2}{8a^4}\Lambda ^2 - \frac{a^4-21 m^4}{128 a^8}\Lambda ^4 + \frac{5 a^6-14 a^4 m^2-123 a^2 m^4+220 m^6}{768 a^{12}}\Lambda ^6 \nn\\[3mm]
&&\quad +\frac{-21 a^8+738 a^6 m^2+3080 a^4 m^4-19266 a^2 m^6+18445 m^8}{32768a^{16}}\Lambda ^8 + \cdots \,.
\eea
The $F_{0,0}$ and $F_{0,1}$ are the same as the known results \cite{Eguchi:2010rf} and \cite{Nishinaka:2011aa}, and the genus one correction $F_{0,2}$ exactly matches the corresponding part \eqref{af022} of the genus one correction computed from the Nekrasov partition function.

\section{$N_F=3$ Model}\label{f3}

In this section, we consider the matrix model, which corresponds to $SU(2)$ gauge theory with $N_F=3$. The action of the matrix model \eqref{po3} is given by
\beq
V(z) = V_0 + \e V_1 = 2\mu_3\log z + 2m_1 \log(z-1) -\frac{\L}{z} + \e\log z \,.
\eeq
The function $f(z)$ reads
\beq
f(z) = \frac{c_1}{z} + \frac{c_2}{z-1} + \frac{c_3}{z^2} \,,
\eeq
where $c_1, c_2$ and $c_3$ are
\bea
c_1 &=& -g_s\sqrt{\b}\sum_{I=1}^N\left\la\frac{2\mu_3 + \e}{\l_I} + \frac{\L}{\l_I^2}\right\ra, \quad c_2 = -g_s\sqrt{\b}\sum_{I=1}^N\left\la\frac{2m_1}{\l_I-1}\right\ra \,,\nn\\[2mm]
c_3 &=& -g_s\sqrt{\b}\sum_{I=1}^N\left\la\frac{\L}{\l_I}\right\ra \,.
\eea
The condition $\la\sum_IV'(\l_I)\ra=0$ yields a constraint
\beq
c_1 + c_2 = 0 \,.
\eeq
From the large $z$ limit of the planar loop equation in \eqref{loopexp} with $W_{0,0}(z)\approx g_s\sqrt{\b}N/z$ and $f(z) \approx (c_2+c_3)/z^2$, we have another constraint
\beq
c_2 + c_3 = m_{\infty}^2 - (\mu_3 + m_1)^2 \,.
\eeq
Since only one parameter among $c_i$'s is independent, we choose $c_3$ to parameterize the spectral curve.

The planar spectral curve for $N_F=3$ model is of the form
\beq
y_0^2 = \frac{P_4(z)}{4z^4(z-1)^2} \,,
\eeq
where the polynomial $P_4(z)$ is
\begin{eqnarray}
 P_4(z) &=& 4m_\infty^2z^4 - 4(B-m_1\Lambda-m_1^2+m_\infty^2)z^3
\nonumber\\[1mm]
&&\quad + (4B-4m_1\Lambda+ \Lambda^2 -4\mu_3\Lambda)z^2 +2\Lambda(2\mu_3-\Lambda)z + \Lambda^2 \,,
\end{eqnarray}
with $B = c_3 -\mu_3\Lambda +\mu_3^2$. Henceforth, we take $\mu_3=m$ and $m_1=m_\infty=0$ for simplicity. Then, $P_4(z)$ reduces to
\begin{eqnarray}
 P_4(z) &=& -4Bz^3 + (4B +\Lambda^2 - 4m\Lambda)z^2 + 2\Lambda(2m-\Lambda)z + \Lambda^2 \,.
\end{eqnarray}

The planar Coulomb branch parameter $a_0$ and the half genus part $a_{0,1}$ are
\bea
a_0 &=& \cint \frac{\sqrt{P_4}}{2z^2(z-1)}dz \,, \label{pa0f3}\\[3mm]
a_{0,1} &=& \cint \frac{\L}{2}\left(\frac{1}{z}-1\right)\frac{dz}{\sqrt{P_4}} = -\frac{1}{2}\frac{\p a_0}{\p m} + \frac{\L}{2}\frac{\p a_0}{\p B} \,,
\eea
with the derivatives of $a_0$,
\beq\label{firstderi}
\frac{\p a_0}{\p m} = \cint \left(-\frac{\L}{z}\right)\frac{dz}{\sqrt{P_4}} \,, \qquad \frac{\p a_0}{\p B} = \cint \left(-\frac{dz}{\sqrt{P_4}}\right) \,.
\eeq
We will need the second derivatives of $a_0$ to compute the genus one part $a_{0,2}$.
\bea
\frac{\p^2 a_0}{\p m^2} &=& \cint 2\L^2(1-z)\frac{dz}{P_4^{3/2}} \label{mm3}\\[1mm]
\frac{\p^2 a_0}{\p m\p B} &=& \cint 2\L z(1-z)\frac{dz}{P_4^{3/2}} \label{mB3}\\[1mm]
\frac{\p^2 a_0}{\p B^2} &=& \cint 2z^2(1-z)\frac{dz}{P_4^{3/2}}  \label{BB3}
\eea

Now, we will find the expression of $a_{0,2}$ in terms of the derivatives of $a_0$. The first term in \eqref{a02} is given by
\beq\label{pft3}
\cint \frac{y_0'^2}{8y_0^3}dz = \cint \frac{1}{16}\left((z^3-z^2)\frac{P_4'^2}{P_4^{5/2}} - 4z(3z-2)\frac{P_4'}{P_4^{3/2}} + \frac{4(3z-2)^2}{z-1}\frac{1}{\sqrt{P_4}}\right)dz \,.
\eeq
The first integral on the right hand side of the above equation can be simplified using the fact
\beq\label{trick}
\frac{P_4'^2}{P_4^{5/2}} = \frac{4}{3}\frac{d^2}{dz^2}\frac{1}{\sqrt{P_4}} +\frac{2}{3}\frac{P_4''}{P_4^{3/2}}  \,,
\eeq
and integrating by parts.
\beq\label{fft3}
\cint \frac{1}{16}(z^3-z^2)\frac{P_4'^2}{P_4^{5/2}}dz = \cint \left(\frac{3z-1}{6}\frac{1}{\sqrt{P_4}} + \frac{1}{24}(z^3-z^2)\frac{P_4''}{P_4^{3/2}}\right)dz \,.
\eeq
The numerator of the last term satisfies the following relation.
\bea
(z^3-z^2)P_4'' &=& 2(z^2-z)P_4' \nn\\[2mm]
&& + 2(4B + \L^2 - 4m\L)(z^2-z^3) + 4\L(2m-\L)(z-z^2)
\eea
Then, we obtain the integral \eqref{fft3} after integration by parts
\bea
\cint \frac{1}{16}(z^3-z^2)\frac{P_4'^2}{P_4^{5/2}}dz &=& \cint \frac{1}{6}(5z-2)\frac{dz}{\sqrt{P_4}} \nn\\[2mm]
&+&  \frac{1}{12}(2m-\L)\frac{\p^2 a_0}{\p m\p B} + \frac{1}{24}(4B+\L^2-4m\L)\frac{\p^2 a_0}{\p B^2} \,,
\eea
via \eqref{mB3} and \eqref{BB3}.
Plugging the above expression in \eqref{pft3}, one has
\bea
\cint \frac{y_0'^2}{8y_0^3}dz &=& \cint \left(\frac{1}{12}\frac{z}{\sqrt{P_4}} + \frac{1}{4(z-1)}\frac{1}{\sqrt{P_4}}\right)dz + \frac{1}{4}\frac{\p a_0}{\p B} \nn\\[3mm]
&& + \frac{1}{12}(2m-\L)\frac{\p^2 a_0}{\p m\p B} + \frac{1}{24}(4B+\L^2-4m\L)\frac{\p^2 a_0}{\p B^2} \,.
\eea
We give the computation of the integrals in the above equation in Appendix \ref{Integral}. With \eqref{result1} and \eqref{result2}, we obtain the first term of $a_{0,2}$ as
\bea\label{ft3}
\cint \frac{y_0'^2}{8y_0^3}dz &=& \frac{1}{12B}a_0 + \frac{1}{12}\frac{\p a_0}{\p B} - \frac{m}{12B}\frac{\p a_0}{\p m} \nn\\[2mm]
&& - \frac{1}{8\L}(4m - \L)\frac{\p^2 a_0}{\p m^2} + \frac{1}{12\L}(-12B + 2m\L - \L^2)\frac{\p^2 a_0}{\p m\p B} \nn\\[2mm]
&& + \frac{1}{24}(4B+\L^2-4m\L)\frac{\p^2 a_0}{\p B^2} \,. \nn\\
\eea
The second term of $a_{0,2}$ in \eqref{a02} takes the form
\bea\label{st3}
\cint \frac{(V_0''+V_0'V_1')^2}{32y_0^3}dz &=& \cint \frac{\L^2}{4}\left((z^3-z^2) + 2(z-z^2) - (1-z)\right)\frac{dz}{P_4^{3/2}} \nn\\[2mm]
&=& -\frac{\L^2}{8}\frac{\p^2 a_0}{\p B^2} + \frac{\L}{4}\frac{\p^2 a_0}{\p m\p B} - \frac{1}{8}\frac{\p^2 a_0}{\p m^2} \,.
\eea
The third term of $a_{0,2}$ is given by
\bea\label{tt3}
\cint\frac{V_1'^2+2V_1''}{8y_0} &=& \cint \frac{1-z}{4\sqrt{P_4}}dz \nn\\[3mm]
&=& - \frac{1}{4B}a_0 + \frac{1}{4}\frac{\p a_0}{\p B} + \frac{m}{4B}\frac{\p a_0}{\p m} \,,
\eea
where we have used the result of \eqref{result1}.
Collecting all three terms \eqref{ft3}, \eqref{st3}, and \eqref{tt3} of $a_{0,2}$, we find that
\bea\label{a02f3}
a_{0,2} &=& -\frac{1}{6B}a_0 + \frac{1}{3}\frac{\p a_0}{\p B} + \frac{m}{6B}\frac{\p a_0}{\p m} \nn\\[2mm]
&& + \frac{1}{4\L}(-2m + \L)\frac{\p^2a_0}{\p m^2} + \frac{1}{6\L}(-6B + m\L - 2\L^2)\frac{\p^2a_0}{\p m\p B} \nn\\[2mm]
&& + \frac{1}{6}(B - m\L + \L^2)\frac{\p^2a_0}{\p B^2} \,.
\eea

The planar and the half genus contribution of the Coulomb branch parameter have been computed, which we reproduce here. \cite{Eguchi:2010rf}, \cite{Nishinaka:2011aa}
\begin{eqnarray}
 a_0 &=& -\sqrt{B}\left(1 + \frac{m}{4B}\Lambda - \frac{B+3m^2}{64B^{2}}\Lambda^2 + \frac{m(5m^2+B)}{256B^3}\Lambda^3\right.
\nonumber\\[2mm]
&& \quad\quad -\frac{3B^2 + 30m^2B + 175m^4}{16384B^4}\Lambda^4 + \frac{m(9B^2 + 70m^2B + 441m^4)}{65536B^5}\Lambda^5 \nn\\[2mm]
&& \quad\quad \left.-\frac{5B^3+105m^2B^2+735m^4B+4851m^6}{1048576B^6}\L^6 + \mathcal{O}(\Lambda^7)\right) \,.\\[4mm]
 a_{0,1} &=& -\frac{1}{8 \sqrt{B}}\Lambda  + \frac{m}{64 B^{3/2}} \Lambda ^2 - \frac{B+3 m^2 }{512 B^{5/2}}\Lambda ^3 + \frac{9 B m+25 m^3}{8192 B^{7/2}}\Lambda ^4
\nonumber\\[3mm]
&&-\frac{9 B^2+90 B m^2+245 m^4}{65536 B^{9/2}} \Lambda^5 +\frac{m \left(75 B^2+490 B m^2+1323 m^4\right)}{524288 B^{11/2}}\Lambda^6 +\cdots.
\nonumber\\
\end{eqnarray}
The genus one part \eqref{a02f3} is given by
\bea
a_{0,2} &=& -\frac{1}{12 \sqrt{B}} + \frac{m}{12 B^{3/2}}\Lambda  +\frac{11 B-51 m^2}{768 B^{5/2}}\Lambda ^2  + \frac{95 m^3-12 B m}{1536 B^{7/2}}\Lambda ^3  \nn\\[3mm]
&& +\frac{111 B^2+1350 B m^2-11725 m^4}{196608 B^{9/2}}\Lambda ^4 -\frac{7\left(15 B^2m+185 B m^3-1638 m^5\right)}{196608 B^{11/2}}\Lambda ^5  \nn\\[3mm]
&& + \frac{1315 B^3+32235 B^2 m^2+381465 B m^4+635481 m^6}{25165824 B^{13/2}}\Lambda ^6 + \cdots\,.
\eea
\\
Inverting the vev $a$ to get $B$, we obtain
\bea
B &=& \left(a^2 -\frac{m\L}{2} + \frac{a^2+m^2}{32a^2}\L^2 + \frac{a^4-6m^2a^2+5m^4}{8192a^6}\L^4 + \frac{5a^4m^2-14a^2m^4+9m^6}{262144a^{10}}\L^6+\cdots\right) \nn\\[4mm]
&& + \e\left(-\frac{\Lambda }{4} -\frac{m }{32 a^2}\Lambda ^2+\frac{ 3 a^2 m-5 m^3}{4096 a^6}\Lambda ^4 - \frac{ 5 a^4 m-28 a^2
   m^3+27 m^5}{262144 a^{10}}\Lambda ^6 + \cdots\right) \nn\\[4mm]
&& + \e^2\left(-\frac{1}{6}+\frac{m \Lambda }{8 a^2} + \frac{a^2-m^2}{64 a^4}\Lambda ^2  +\frac{5 m^3-3 a^2 m}{256 a^6}\Lambda ^3 -\frac{2 a^4-25 a^2 m^2+21 m^4}{8192 a^8}\Lambda ^4 \right. \nn\\[3mm]
&& \hskip1cm +\frac{35 a^4 m-210 a^2 m^3+207 m^5}{65536 a^{10}}\Lambda ^5 \nn\\[3mm]
&& \hskip1cm + \left.\frac{215 a^6+5719 a^4
   m^2+75477 a^2 m^4+691933 m^6}{4194304 a^{12}}\Lambda ^6 + \cdots\right) \,.
\eea
\\
The free energy can be computed from the derivative of $F$ with respect to $\L$.
\beq
\L\frac{\p}{\p \L}F = -g_s\sqrt{\b}\sum_I\left\la\frac{\L}{\l_I}\right\ra = c_3 = B + m\L - m^2 \,.
\eeq
The planar and the half genus part of the free energy are integrated to
\bea
F_{0,0} &=& (a^2-m^2)\log\Lambda+\frac{m\Lambda}{2} + \frac{a^2 + m^2}{64a^2} \Lambda^2
\nonumber\\[3mm]
&& + \frac{a^4 -6m^2a^2 + 5m^4}{32768\,a^6}\Lambda^4 +\frac{5 a^4 m^2-14 a^2 m^4+9 m^6}{1572864\, a^{10}}\Lambda^6 + \cdots \,. \\[4mm]
F_{0,1} &=&  - \frac{\Lambda}{4} - \frac{m}{64a^2}\Lambda^2 + \frac{3a^2m-5m^3}{16384 \,a^6}\Lambda^4  -\frac{5 a^4 m-28 a^2 m^3+27 m^5}{1572864\, a^{10}}\Lambda^6
+ \cdots.
\eea
The genus one part is given by
\bea
F_{0,2} &=& -\frac{1}{6}\log\L +\frac{m}{8 a^2}\L + \frac{a^2-m^2}{128 a^4}\Lambda ^2  +\frac{5 m^3-3 a^2 m}{768 a^6}\Lambda ^3 -\frac{2 a^4-25 a^2 m^2+21 m^4}{32768a^8}\Lambda ^4 \nn\\[3mm]
&& + \frac{35 a^4 m-210 a^2 m^3+207 m^5}{327680a^{10}}\Lambda ^5  \nn\\[3mm]
&& + \frac{215 a^6+5719 a^4 m^2+75477 a^2 m^4+691933
   m^6}{25165824 a^{12}}\Lambda ^6 + \cdots \,.
\eea
Again, the free energy of the matrix model agrees with that of the gauge theory computed from the Nekrasov partition function in Appendix \ref{af3}.

\section{$N_F=4$ Model}\label{f4}

In this section, we consider the matrix model with action \eqref{po4}, which is related to the superconformal case of $SU(2)$ gauge theory with four flavors. The Penner type potential reads
\bea
V_0(z) &=& 2m_0\log z + 2m_1\log(z-1) + 2m_2\log(z-q) \,,\nn\\[1mm]
V_1(z) &=& \log z \,.
\eea
We evaluate the function $f(z)$ in \eqref{f} as
\beq
f(z)= \frac{c_0}{z} + \frac{c_1}{z-1} + \frac{c_2}{z-q} \,,
\eeq
where the $c_i$'s are given by
\bea
c_0 &=& -g_s\sqrt{\b}\sum_I\left\la\frac{2m_0+\e}{\l_I}\right\ra, \quad\quad c_1 = -g_s\sqrt{\b}\sum_I\left\la\frac{2m_I}{\l_I-1}\right\ra \,,\nn\\[2mm]
c_2 &=& -g_s\sqrt{\b}\sum_I\left\la\frac{2m_2}{\l_I - q}\right\ra \,.
\eea
From the equation of motion $\la\sum_I V'(\l_I)\ra=0$, it follows that
\beq
\sum_{i=0}^2 c_i = 0 \,.
\eeq
And from the asymtotic behavior for $z\rightarrow\infty$ of the leading order loop equation in \eqref{loopexp}, one finds that the parameters satisfy
\beq\label{constraint4}
c_1 + qc_2 = m_\infty^2 - \left(\sum_{i=0}^2m_i \right)^2 \,.
\eeq
Therefore, we have a single parameter left to describe the spectral curve, which we take to be $c_0$.

The planar spectral curve is
\beq
y_0^2 = \frac{P_4(z)}{z^2(z-1)^2(z-q)^2} \,,
\eeq
with $P_4(z)$ a polynomial of degree four. We will consider the case where the mass $\mu_i$ of all four hypermutiplets is equal to $m$ such that the mass parameters $m_i$ are set to $m_0=m_\infty=0$ and $m_1=m_2=m$. Then, the polynomial $P_4(z)$ becomes
\beq
P_4(z) = Cz^3 + ((1-q)^2m^2 - C(1+q))z^2 + Cqz \,,
\eeq
with $C \equiv qc_0$.

In this case, the Coulomb branch parameters up to half genus are
\bea
a_0 &=& \cint \frac{1}{z(z-1)(z-q)}\sqrt{P_4}\,dz \,,\\[4mm]
a_{0,1} &=& \cint \left(-\frac{m(1+q)}{2}\frac{1}{\sqrt{P_4}} - \frac{m(1-q)^2}{2}\frac{z}{(z-1)(z-q)}\frac{dz}{\sqrt{P_4}}\right) \nn\\[3mm]
&=& -m(1+q)\frac{\p a_0}{\p C} - \frac{1}{2}\frac{\p a_0}{\p m} \,,
\eea
where we have utilized in the last equality
\bea
\frac{\p a_0}{\p C} &=& \cint\frac{1}{2}\frac{dz}{\sqrt{P_4}} \,,\label{C4}\\[2mm]
\frac{\p a_0}{\p m} &=& \cint m(1-q)^2\frac{z}{(z-1)(z-q)}\frac{dz}{\sqrt{P_4}} \,.\label{m4}
\eea
We compute the second derivatives of $a_0$ as
\bea
\frac{\p^2a_0}{\p C\p m} &=& \cint \frac{-m(1-q)^2}{2}z^2\frac{dz}{P_4^{3/2}} \,,\label{Cm4}\\[3mm]
\frac{\p^2a_0}{\p m^2} &=& \cint \left(\frac{-m^2(1-q)^4z^3}{(z-1)(z-q)}\frac{1}{P_4^{3/2}}+ \frac{(1-q)^2z}{(z-1)(z-q)}\frac{1}{\sqrt{P_4}}\right)dz \,.\label{mm4}
\eea

Now, we calculate the genus one part $a_{0,2}$ in terms of the derivatives of $a_0$. The first term of $a_{0,2}$ in \eqref{a02} becomes
\bea\label{pft4}
\cint \frac{y_0'^2}{8y_0^3}dz &=& \cint \left(\frac{z(z-1)(z-q)}{32}\frac{P_4'^2}{P_4^{5/2}} - \frac{3z^2-2(1+q)z+q}{8}\frac{P_4'}{P_4^{3/2}}\right. \nn\\[3mm]
&&\hskip2cm \left. + \frac{(3z^2-2(1+q)z+q)^2}{8z(z-1)(z-q)}\frac{1}{\sqrt{P_4}}\right)dz \,.
\eea
Using \eqref{trick} and integrating by parts, the first integral on the right hand side can be written as
\bea\label{midresult1}
\cint \left(\frac{3z-(1+q)}{12}\frac{1}{\sqrt{P_4}} + \frac{z(z-1)(z-q)}{48}\frac{P_4''}{P_4^{3/2}}\right)dz \,.
\eea
This can be written using \eqref{result14} as
\bea\label{fft4}
&& \cint \left(\frac{3}{8}\frac{z}{\sqrt{P_4}} + \frac{m^2(1-q)^2q}{8}\frac{z}{P_4^{3/2}}\right)dz - \frac{1}{4}(1+q)\frac{\p a_0}{\p C} -\frac{m^2(1-q)^2}{6C} \frac{\p a_0}{\p C} \nn\\[3mm]
&& - \frac{m(m^2(1-q)^2-C(1+q))}{6C}\frac{\p^2 a_0}{\p C\p m}\,.
\eea
The second term in \eqref{pft4} is integrated by parts to
\beq\label{fst4}
\cint \left(-\frac{3}{2}z + \frac{1}{2}(1+q)\right)\frac{dz}{\sqrt{P_4}} =  \cint \frac{-3}{2}\frac{z}{\sqrt{P_4}}dz + (1+q)\frac{\p a_0}{\p C} \,.
\eeq
The third term in \eqref{pft4} can be written as
\bea\label{ftt4}
&&\cint \frac{1}{8}\left(9z -3(1+q) + \frac{q}{z} + (1-q)^2\frac{z}{(z-1)(z-q)}\right)\frac{dz}{\sqrt{P_4}} \nn\\[3mm]
&&\qquad = \cint \frac{9}{8}\frac{z}{\sqrt{P_4}}dz -\cint \frac{m^2(1-q)^2-C(1+q)}{8}\frac{qz}{P_4^{3/2}}dz  \nn\\[3mm]
&&\qquad\quad - \frac{3}{4}(1+q)\frac{\p a_0}{\p C} + \frac{Cq}{2m(1-q)^2}\frac{\p^2 a_0}{\p C\p m} + \frac{1}{8m}\frac{\p a_0}{\p m} \,,
\eea
where \eqref{result44} has been used. Adding \eqref{fft4}, \eqref{fst4}, and \eqref{ftt4}, the first term of $a_{0,2}$ in \eqref{pft4} becomes
\bea
\cint \frac{y_0'^2}{8y_0^3}dz &=& \cint \frac{Cq(1+q)}{8}\frac{z}{P_4^{3/2}}dz \nn\\[3mm]
&& - \frac{m^2(1-q)^2}{6C}\frac{\p a_0}{\p C} - \frac{m(m^2(1-q)^2-C(1+q))}{6C}\frac{\p^2 a_0}{\p C\p m} \nn\\[3mm]
&& + \frac{Cq}{2m(1-q)^2}\frac{\p^2 a_0}{\p C\p m} + \frac{1}{8m}\frac{\p a_0}{\p m} \,.
\eea
For the integration of $z/P_4^{3/2}$ in the above equation, we use \eqref{result24}, which results in
\bea\label{ft4}
\cint \frac{y_0'^2}{8y_0^3}dz &=& \frac{1}{8}(1+q)\frac{\p a_0}{\p C} + \frac{1+q}{m(1-q)^2}(m^2(1-q)^2-C(1+q))\frac{\p^2a_0}{\p C\p m} \nn\\[3mm]
&& - \frac{m^2(1-q)^2}{6C}\frac{\p a_0}{\p C} - \frac{m(m^2(1-q)^2-C(1+q))}{6C}\frac{\p^2 a_0}{\p C\p m} \nn\\[3mm]
&& + \frac{Cq}{2m(1-q)^2}\frac{\p^2 a_0}{\p C\p m} + \frac{1}{8m}\frac{\p a_0}{\p m} \,.
\eea

Next, we compute the second term of $a_{0,2}$ in \eqref{a02}.
\bea\label{st4}
&&\cint \frac{(V_0''+V_0'V_1')^2}{32y_0^3}dz \nn\\[3mm]
&&\quad = \cint \frac{m^2}{8}\left(\frac{(1-q)^4z^3}{(z-1)(z-q)} + (1+q)^2z^3 + (1+q)^3z^2 - 8q(1+q)z^2 + q(1+q)^2z\right)\frac{dz}{P_4^{3/2}} \nn\\[3mm]
&&\quad = -\frac{1}{8}\frac{\p^2a_0}{\p m^2} + \frac{1}{8m}\frac{\p a_0}{\p m} \nn\\[3mm]
&&\quad\quad + \cint \frac{m^2}{8}\left(\frac{(1+q)^2}{C}P_4 - \frac{(1+q)^2}{C}(m^2(1-q)^2-2C(1+q))z^2 -8q(1+q)z^2\right)\frac{dz}{P_4^{3/2}} \nn\\[3mm]
&&\quad = -\frac{1}{8}\frac{\p^2a_0}{\p m^2} + \frac{1}{8m}\frac{\p a_0}{\p m} + \frac{m^2(1+q)^2}{4C}\frac{\p a_0}{\p C} \nn\\[3mm]
&&\quad\quad + \frac{m}{4C}\frac{(1+q)^2}{(1-q)^2}(m^2(1-q)^2-2C(1+q))\frac{\p^2a_0}{\p C\p m} + 2m\frac{q(1+q)}{(1-q)^2}\frac{\p^2a_0}{\p C\p m} \,.
\eea

The third term of the genus one part of the Coulomb branch parameter \eqref{a02} is calculated to
\bea\label{tt4}
&&\cint \frac{V_1'^2+2V_1''}{8y_0}dz \nn\\[3mm]
&&\hskip1.5cm = \cint \frac{-1}{8}\left(z-(1+q)+\frac{q}{z}\right)\frac{dz}{\sqrt{P_4}} \nn\\[3mm]
&&\hskip1.5cm = \frac{m^2(1-q)^2-C(1+q)}{8C}\frac{\p a_0}{\p C} + \frac{(m^2(1-q)^2-C(1+q))^2}{8Cm(1-q)^2}\frac{\p^2a_0}{\p C\p m} \nn\\[3mm]
&&\hskip1.5cm\qquad - \frac{Cq}{2m(1-q)^2}\frac{\p^2a_0}{\p C\p m} + \frac{1}{4}(1+q)\frac{\p a_0}{\p C} - \frac{Cq}{2m(1-q)^2}\frac{\p^2a_0}{\p C\p m} \nn\\[3mm]
&&\hskip1.5cm\qquad +\cint \frac{q}{8}(m^2(1-q)^2-C(1+q))\frac{z}{P_4^{3/2}}dz \nn\\[3mm]
&&\hskip1.5cm = \frac{m^2(1-q)^2}{4C}\frac{\p a_0}{\p C} + \frac{(m^2(1-q)^2-C(1+q))^2}{4Cm(1-q)^2}\frac{\p^2a_0}{\p C\p m} \nn\\[3mm]
&&\hskip1.5cm\qquad - \frac{Cq}{m(1-q)^2}\frac{\p^2a_0}{\p C\p m} \,,
\eea
where \eqref{result34}, \eqref{C4}, and \eqref{result44} have been used in the second equality and \eqref{result24} was used in the last equality.
From the results of the three terms of $a_{0,2}$ in \eqref{ft4}, \eqref{st4}, and \eqref{tt4}, we find the genus one part of the Coulomb branch parameter \eqref{a02}
\bea\label{a02fl4}
a_{0,2} &=& \frac{1}{8}(1+q)\frac{\p a_0}{\p C} + \frac{m^2(1-q)^2}{12C}\frac{\p a_0}{\p C} - \frac{Cq}{2m(1-q)^2}\frac{\p^2a_0}{\p C\p m} \nn\\[3mm]
&& + \frac{1}{8m}\frac{1+q}{(1-q)^2}(m^2(1-q)^2-C(1+q))\frac{\p^2a_0}{\p C\p m} - \frac{m}{6C}(m^2(1-q)^2-C(1+q))\frac{\p^2a_0}{\p C\p m} \nn\\[3mm]
&& + \frac{1}{8}\frac{\p^2 a_0}{\p m^2} - \frac{m^2(1+q)^2}{4C}\frac{\p a_0}{\p C} - \frac{m}{4C}\frac{(1+q)^2}{(1-q)^2}(m^2(1-q)^2-C(1+q))\frac{\p^2 a_0}{\p C\p m} \nn\\[3mm]
&& + \frac{m}{4}\frac{(1+q)^3}{(1-q)^2}\frac{\p^2a_0}{\p C\p m} - 2m\frac{q(1+q)}{(1-q)^2}\frac{\p^2a_0}{\p C\p m} \nn\\[3mm]
&& + \frac{(m^2(1-q)^2-C(1+q))^2}{4Cm(1-q)^2}\frac{\p^2a_0}{\p C\p m} \,.
\eea
\\

We give a few leading order terms in $m$ of the planar and the half genus contribution of the Coulomb branch parameter. \cite{Eguchi:2010rf}, \cite{Nishinaka:2011aa}
\bea\label{a04}
a_0 &=& i\sqrt{C}\left(h_0(q) - h_1(q)\frac{m^2}{C} - \frac{h_2(q)}{3}\frac{m^4}{C^2} - \frac{h_3(q)}{5}\frac{m^6}{C^3} -\frac{h_4(q)}{7} \frac{m^8}{C^4}+\mathcal{O}\left(\frac{m^{10}}{C^5}\right)\right),\label{eq:a0_NF=4}
\nonumber\\[4mm]
 a_{0,1} &=& i\left(g_1(q)\frac{m}{\sqrt{C}} + g_2(q) \frac{m^3}{C^{3/2}} + g_3(q)\frac{m^5}{C^{5/2}} + g_4(q)\frac{m^7}{C^{7/2}}+\cdots\right) \,,
\eea
where the functions $h_i(q)$'s are obtained from the expansion of the hypergeometric function,
\bea
h_0(q) &=& 1+\frac{1}{4}q + \frac{9}{64}q^2 + \frac{25}{256}q^3 + \frac{1225}{16384}q^4 + \mathcal{O}(q^5),
\\[2mm]
h_1(q) &=& \frac{1}{2}+\frac{1}{8}q + \frac{1}{128}q^2 +\frac{1}{512}q^3 + \frac{25}{32768}q^4 + \mathcal{O}(q^5),
\\[2mm]
h_2(q) &=& \frac{3}{8} + \frac{27}{32}q + \frac{27}{512}q^2 + \frac{3}{2048}q^3 + \frac{27}{131072}q^4 + \mathcal{O}(q^5),
\\[2mm]
h_3(q) &=& \frac{5}{16}+\frac{125}{64}q + \frac{1125}{1024}q^2 + \frac{125}{4096}q^3 + \frac{125}{262144}q^4 + \mathcal{O}(q^5), \\[2mm]
h_4(q) &=& \frac{35}{128} + \frac{1715}{512}q + \frac{42875}{8192}q^2 + \frac{42875}{32768}q^3 + \frac{42875}{2097152}q^4 + \mathcal{O}(q^5)
\eea
and the function $g_i(q)$ is given in terms of $h_i(q)$'s by
\bea
 g_1(q) &=& \frac{2h_1(q) - (1+q)h_0(q)}{2},
\\[1mm]
 g_2(q) &=& \frac{4h_2(q)-3(1+q)h_1(q)}{6},
\\[1mm]
 g_3(q) &=& \frac{6h_3(q) - 5(1+q)h_2(q)}{10},
\\[1mm]
 g_4(q) &=& \frac{8h_4(q)-7(1+q)h_3(q)}{14}.
\eea
The genus one part of the vev \eqref{a02fl4} is computed to
\bea
a_{0,2} &=& i\left(k_1(q)\frac{1}{\sqrt{C}} + k_2(q)\frac{m^2}{C^{3/2}} + k_3(q)\frac{m^4}{C^{5/2}} + k_4(q)\frac{m^6}{C^{7/2}} + \cdots\right) \,,
\eea
where $k_i(q)$ is related to $h_i(q)$'s as
\bea
k_1(q) &=& -\frac{1}{16}(-(q+1) h_0(q) + 2h_1(q)) \,,\\
k_2(q) &=& -\frac{1}{48}\left(4(q^2+4q+1)h_0(q) - 17(q+1)h_1(q) + 12h_2(q)\right) \,,\\
k_3(q) &=& -\frac{1}{48}\left(12(q^2 + 4q + 1)h_1(q) - 31(q+1)h_2(q) + 18h_3(q)\right) \,,\\
k_4(q) &=& -\frac{1}{48}\left(20(q^2+4q+1)h_2(q) - 45(q+1)h_3(q) + 24h_4(q)\right) \,.
\eea
With the explicit results of the Coulomb branch parameter, we solve the equation $a = a_0 + \e a_{0,1} + \e^2 a_{0,2}$ for $C$.
\bea
C &=& -a^2\left(\frac{1}{h_0(q)^2} - \frac{2h_1(q)}{h_0(q)}\frac{m^2}{a^2} + \frac{2h_0(q)h_2(q) - 3h_1(q)^2}{3}\frac{m^4}{a^4}\right.
\nonumber \\[3mm]
&& \qquad\quad \left. - \frac{10h_0(q)h_1(q)^3 -10h_0(q)^2h_1(q)h_2(q) + 2h_0(q)^3h_3(q)}{5}\frac{m^6}{a^6} + \cdots\right)
\nonumber\\[4mm]
&& + \epsilon\, m\left(\frac{2(1+q)h_0(q)-4h_1(q)}{h_0(q)} + \frac{8h_0(q)h_2(q)-12h_1(q)^2}{3}\frac{m^2}{a^2} \right.
\nonumber\\[3mm]
&&\qquad\quad\left.- \frac{12h_0(q)\{5h_1(q)^3 - 5h_0(q)h_1(q)h_2(q) + h_0(q)^2h_3(q)\}}{5}\frac{m^4}{a^4} + \cdots
\right)
\nonumber\\[4mm]
&& + \e^2 \left(-\frac{(1+q)h_0(q) - 2 h_1(q)}{8 h_0(q)} \right.\nn\\[3mm]
&&\qquad\quad + \frac{15 h_1(q)^2 + h_0(q)^2 (-1 + q)^2 - h_0(q) (6 h_2(q) + 5 h_1(q) (1 + q))}{12}\frac{m^2}{a^2} \nn\\[3mm]
&&\qquad\quad + \frac{h_0(q)}{12}\{81 h_1(q)^3 - 3 h_0(q) h_1(q)(23 h_2(q) + 5(1+q) h_1(q)) \nn\\[3mm]
&&\qquad\qquad\qquad \left. + h_0(q)^2 (9 h_3(q) + 10 (1+q)h_2(q))\}\frac{m^4}{a^4} + \cdots \right)\nn\\
\eea

Now, we can compute the free energy by
\beq
\frac{\p}{\p q}F = -g_s\sqrt{\b}\left\la\sum_I \frac{2m}{\l_I-q}\right\ra = c_2 \,.
\eeq
From \eqref{constraint4}, $c_2$ is related to $C \equiv qc_0$, thus we obtain
\beq
\frac{\p}{\p q}F = \frac{1}{1-q}\left(4m^2 - \frac{C}{q}\right) \,.
\eeq
The $F_{0,0}$ and $F_{0,1}$ are given by
\bea
F_{0,0} &=& (a^2-m^2)\log q + \frac{a^4 +6a^2 m^2 + m^4}{2a^2}q
\nonumber\\[3mm]
&& + \frac{\left(13 a^8+100 a^6 m^2+22 a^4 m^4-12 a^2 m^6+5 m^8\right)}{64 a^6}q^2
\nonumber\\[3mm]
&& + \frac{23 a^{12}+204 a^{10} m^2+51 a^8 m^4-48 a^6 m^6+45 a^4 m^8-28 a^2 m^{10}+9 m^{12}}{192 a^{10}}q^3
\nonumber\\[3mm]
&& + \frac{1}{32768 a^{14}}\left(2701 a^{16}+26440 a^{14} m^2+7164 a^{12} m^4-9000 a^{10} m^6\right.
\nonumber\\[3mm]
&&\qquad \left.+12190 a^8 m^8-13384 a^6 m^{10}+10908 a^4 m^{12}-5720 a^2 m^{14}+1469 m^{16}\right)q^4 + \cdots \,,
\nonumber\\
\\[4mm]
F_{0,1} &=& -\frac{2m(a^2+m^2)}{2a^2}q - \frac{9 a^6 m+11 a^4 m^3-9 a^2 m^5+5 m^7 }{16 a^6}q^2
\nonumber\\[3mm]
&&-\frac{38 a^{10} m+51 a^8 m^3-72 a^6 m^5+90 a^4 m^7-70 a^2 m^9+27 m^{11}}{96 a^{10}}q^3
\nonumber\\[3mm]
&&-\frac{1}{4096 a^{14}} \left(1257 a^{14}m+1791 a^{12} m^3-3375 a^{10} m^5+6095 a^8 m^7-8365 a^6 m^9\right.
\nonumber\\[3mm]
&&\qquad\qquad\left. +8181 a^4 m^{11}-5005 a^2 m^{13}+1469 m^{15}\right)q^4\;+ \cdots.
\eea
\\
Finally, we obtain the genus one free energy $F_{0,2}$.
\bea
F_{0,2} &=& \frac{a^4+6 a^2 m^2+m^4}{8 a^4}q + \frac{9 a^8+64 a^6 m^2-70 a^4 m^4+40 a^2 m^6+21 m^8}{128 a^8}q^2 \nn\\[3mm]
&& + \frac{19 a^{12}+147 a^{10} m^2-300 a^8 m^4+470 a^6 m^6-357 a^4 m^8+39 a^2
   m^{10}+110 m^{12}}{384 a^{12}}q^3 \nn\\[3mm]
&& + \frac{1}{32768 a^{16}}\left(1257 a^{16}+10276 a^{14} m^2-28776 a^{12} m^4+67460 a^{10} m^6-105630a^8 m^8 \right. \nn\\[3mm]
&&\qquad\qquad  \left. +103308 a^6 m^{10}-43120 a^4 m^{12}-15028 a^2 m^{14}+18445
   m^{16}\right)q^4 +\cdots\,. \nn\\
\eea
The genus one part is identical to \eqref{af024}, which is calculated in $\CN=2$ $SU(2)$ superconformal gauge theory with four flavors.

\newpage
\section{Summary and Discussion}\label{sum}

In this paper, we have investigated the $\b$-deformed matrix models with logarithmic potentials, which have been suggested to explain the AGT relation. Specifically, we have computed the genus one part of the free energy in the matrix models, which describe $\CN=2$ $SU(2)$ gauge theories with $N_F=2,3,$ and 4 flavors. We have checked that the results obtained from the matrix model nicely match the free energies computed from the Nekrasov partition function.

It would be interesting to generalize the duality to $SU(N_c)$ gauge group, in which case a multi-matrix model should be considered. \cite{Itoyama:2009sc} \cite{Schiappa:2009cc} One can also consider extending the results to higher genus parts of the free energy. \cite{Chekhov:2010zg}, \cite{Chekhov:2006rq}, \cite{Brini:2010fc} The $\b$-deformed matrix model is also related to the refinement of the topological B-model. \cite{Aganagic:2011mi} The study of matrix model along this line would be highly interesting. \\

\acknowledgments

We thank Seungjoon Hyun for discussions.
This research was supported by the National Research Foundation (NRF) of Korea grant funded by the Korea government(MEST) with the grant number 2012046278 and by Basic Science Research Program through the NRF of Korea funded by the MEST (2012R1A1A2004410).\\

\appendix

\section{Calculation of Integrals}\label{Integral}

In this appendix, we show the details of calculations of the results used throughout the paper.

\subsection{Integrals for $N_F=3$ Model}

Consider the $a_0$ in \eqref{pa0f3} in the form of
\beq\label{a0app}
a_0 = \cint \frac{1}{2}\left(\frac{1}{z-1} - \frac{1}{z} - \frac{1}{z^2}\right)\sqrt{P_4}\,dz \,.
\eeq
We will rewrite the three terms in the above equation in turn.
Through the relation
\beq
\frac{1}{z-1} = (-4Bz^2 + (-4m\L + + \L^2)z - \L^2)\frac{1}{P_4} \,,
\eeq
the first integral is written as
\bea\label{a0ft}
\cint \frac{1}{z-1}\sqrt{P_4}\,dz &=& \cint (-4Bz^2 + (-4m\L + \L^2)z - \L^2)\frac{dz}{\sqrt{P_4}} \nn\\[2mm]
&=& \cint \left(-\frac{1}{3}(8B+4m\L-\L^2)z - \frac{1}{3}(4m\L + \L^2)\right)\frac{dz}{\sqrt{P_4}} \,,
\eea
where the following condition has been used to eliminate $z^2$ term.
\bea\label{midcal1}
 0 &=& \cint \frac{d}{dz}\sqrt{P_4}\,dz = \cint \frac{P_4'}{2\sqrt{P_4}}dz \nn\\[2mm]
&=& \cint (-6Bz^2 + (4B-4m\L+\L^2)z + \L(2m-\L))\frac{dz}{\sqrt{P_4}}
\eea
Similar manipulations give the second term in \eqref{a0app}.
\bea\label{a0st}
\cint \frac{1}{z}\sqrt{P_4}\,dz &=& \cint \left(-4Bz^2 + (4B-4m\L+\L^2)z + 2\L(2m-\L) + \frac{\L^2}{z}\right)\frac{dz}{\sqrt{P_4}} \nn\\[2mm]
&=& \cint \left(\frac{1}{3}(4B-4m\L+\L^2) + \frac{4}{3}\L(2m-\L) + \frac{\L^2}{z}\right)\frac{dz}{\sqrt{P_4}}
\eea
The third term of \eqref{a0app} can be written as
\bea\label{a0tt}
\cint \frac{1}{z^2}\sqrt{P_4}\,dz &=& \cint \frac{1}{z}\left(\frac{P_4'}{2\sqrt{P_4}}\right)dz \nn\\[2mm]
&=& \cint\left(-6Bz + (4B-4m\L+\L^2) + \frac{\L(2m-\L)}{z}\right)\frac{dz}{\sqrt{P_4}} \,.
\eea
Adding \eqref{a0ft}, \eqref{a0st}, and \eqref{a0tt}, one finds \eqref{a0app} to be
\bea\label{result1}
a_0 = \cint Bz\frac{dz}{\sqrt{P_4}} + 2B\frac{\p a_0}{\p B} + m\frac{\p a_0}{\p m} \,,
\eea
where \eqref{firstderi} has been used.
\\

Next, we rewrite the following integral
\bea\label{midcal2}
\cint \frac{1}{z-1}\frac{dz}{\sqrt{P_4}} &=& \cint (-4Bz^2 + (-4m\L+\L^2)z - \L^2)\frac{dz}{P_4^{3/2}} \nn\\[2mm]
&=& \frac{2B}{\L}\frac{\p^2 a_0}{\p m\p B} + \cint (-(4B+4m\L-\L^2)z-\L^2)\frac{dz}{P_4^{3/2}}   \nn\\[2mm]
&=& \frac{2B}{\L}\frac{\p^2a_0}{\p m\p B} + \frac{1}{2\L^2}(4B+4m\L-\L^2)\frac{\p^2a_0}{\p m^2} \nn\\[2mm]
&& +  \cint (-4B-4m\L)\frac{dz}{P_4^{3/2}}\,,
\eea
where \eqref{mB3} and \eqref{mm3} have been used in the second and third equality. The integral in the last line can be obtained from the relation
\bea
0 &=& \cint \frac{P_4'}{P_4^{3/2}}dz \nn\\[2mm]
&=& \cint (-12Bz^2 + 2(4B-4m\L+\L^2)z + 4m\L - 2\L^2)\frac{dz}{P_4^{3/2}} \nn\\[2mm]
&=& \frac{6B}{\L}\frac{\p^2a_0}{\p m\p B} + \frac{1}{\L^2}(2B+4m\L-\L^2)\frac{\p^2a_0}{\p m^2} \nn\\[2mm]
&&  + \cint (-4B-4m\L)\frac{dz}{P_4^{3/2}} \,.
\eea
Thus, we find the integral \eqref{midcal2} to be
\beq\label{result2}
\cint \frac{1}{z-1}\frac{dz}{\sqrt{P_4}} = -\frac{4B}{\L}\frac{\p^2a_0}{\p m\p B} + \frac{1}{2\L^2}(-4m\L + \L^2)\frac{\p^2a_0}{\p m^2} \,.
\eeq
\\

\subsection{Integrals for $N_F=4$ Model}

First, we consider the integral in \eqref{midresult1}
\bea\label{midcal4}
&& \cint z(z-1)(z-q)\frac{P_4''}{P_4^{3/2}}dz \nn\\[2mm]
&&\quad = \cint \frac{1}{C}(P_4 - m^2(1-q)^2z^2)\frac{P_4''}{P_4^{3/2}}dz \nn\\[2mm]
&&\quad = \cint \left(\frac{6Cz-2C(1+q)+2m^2(1-q)^2}{C\sqrt{P_4}} - \frac{m^2(1-q)^2}{C}\frac{z^2P_4''}{P_4^{3/2}}\right)dz \,.\nn\\
\eea
We rewrite the second term in the last line via
\beq
z^2P_4'' = 6P_4-4(m^2(1-q)^2 - C(1+q))z^2 - 6Cqz \,,
\eeq
and \eqref{Cm4} to obtain
\bea
\cint \frac{-m^2(1-q)^2}{C}\frac{z^2P_4''}{P_4^{3/2}}dz &=& \cint \left(\frac{-6m^2(1-q)^2}{C}\frac{1}{\sqrt{P_4}} + 6m^2(1-q)^2q\frac{z}{P_4^{3/2}}\right)dz \nn\\[3mm]
&&\quad - \frac{8m}{C}(m^2(1-q)^2-C(1+q))\frac{\p^2a_0}{\p C\p m} \,.
\eea
Substituting the above equation into \eqref{midcal4}, we get
\bea\label{result14}
&&\cint z(z-1)(z-q)\frac{P_4''}{P_4^{3/2}}dz \nn\\[2mm]
&&\quad = \cint \left(6z-2(1+q)-\frac{4m^2(1-q)^2}{C}\right)\frac{dz}{\sqrt{P_4}} +\cint 6m^2(1-q)^2q\frac{z}{P_4^{3/2}}dz \nn\\[2mm]
&&\qquad - \frac{8m}{C}(m^2(1-q)^2-C(1+q))\frac{\p^2a_0}{\p C\p m} \nn\\[3mm]
&&\quad = \cint\left(\frac{6z}{\sqrt{P_4}} + 6m^2(1-q)^2q\frac{z}{P_4^{3/2}}\right)dz - \left(4(1+q)+\frac{8m^2(1-q)^2}{C}\right)\frac{\p a_0}{\p C} \nn\\[2mm]
&&\qquad - \frac{8m}{C}(m^2(1-q)^2-C(1+q))\frac{\p^2a_0}{\p C\p m} \,,
\eea
where \eqref{C4} has been used in the last equality. In the above expression, we need to compute the integrals of $z/\sqrt{P_4}$ and $z/P_4^{3/2}$, which we do in the following.

So, we rewrite $\p a_0/\p C$ of \eqref{C4} to obtain the integral of $z/P_4^{3/2}$ through
\beq
zP_4' = 3P_4 - (m^2(1-q)^2-C(1+q))z^2-2Cqz\,,
\eeq
such that
\bea
\frac{\p a_0}{\p C} &=& \cint \frac{dz}{2\sqrt{P_4}} = \cint \frac{z}{4}\frac{P_4'}{P_4^{3/2}}dz \nn\\[3mm]
&=& \cint \frac{1}{4}\big(3P_4 - (m^2(1-q)^2 - C(1+q))z^2 - 2Cqz\big)\frac{dz}{P_4^{3/2}} \nn\\[3mm]
&=& \frac{3}{2}\frac{\p a_0}{\p C} + \frac{m^2(1-q)^2-C(1+q)}{2m(1-q)^2}\frac{\p^2a_0}{\p C\p m} + \cint \frac{-Cqz}{2}\frac{dz}{P_4^{3/2}} \,.
\eea
Therefore, we obtain
\beq\label{result24}
\cint \frac{Cqz}{P_4^{3/2}}dz = \frac{\p a_0}{\p C} + \frac{m^2(1-q)^2-C(1+q)}{m(1-q)^2}\frac{\p^2a_0}{\p C\p m} \,.
\eeq

Next, we will compute the integral of $z/\sqrt{P_4}$ using the relation
\bea
z^2P_4' &=& 3zP_4 - \frac{m^2(1-q)^2-C(1+q)}{C}P_4  -2Cqz^2\nn\\[2mm]
&&\quad + \frac{(m^2(1-q)^2-C(1+q))^2}{C}z^2 + (m^2(1-q)^2-C(1+q))qz \,.
\eea
It follows that
\bea
\cint \frac{z}{\sqrt{P_4}}dz &=& \cint \frac{z^2}{4}\frac{P_4'}{P_4^{3/2}} \nn\\[2mm]
&=& \cint \left(\frac{3z}{4\sqrt{P_4}}-\frac{m^2(1-q)^2-C(1+q)}{4C}\frac{1}{\sqrt{P_4}}-\frac{Cqz^2}{2P_4^{3/2}}\right.\nn\\[3mm]
&&\qquad + \left.\frac{(m^2(1-q)^2-C(1+q))^2}{4C}\frac{z^2}{P_4^{3/2}} + \frac{m^2(1-q)^2-C(1+q)}{4}\frac{qz}{P_4^{3/2}}\right)dz \nn\\[3mm]
&=& \cint \frac{3}{4}\frac{z}{\sqrt{P_4}}dz -\frac{m^2(1-q)^2-C(1+q)}{2C}\frac{\p a_0}{\p C} + \frac{Cq}{m(1-q)^2}\frac{\p^2a_0}{\p C\p m} \nn\\[3mm]
&&\qquad - \frac{(m^2(1-q)^2-C(1+q))^2}{2Cm(1-q)^2}\frac{\p^2a_0}{\p C\p m} \nn\\[3mm]
&&\qquad + \frac{m^2(1-q)^2-C(1+q)}{4C}\left(\frac{\p a_0}{\p C} + \frac{m^2(1-q)^2-C(1+q)}{m(1-q)^2}\frac{\p^2a_0}{\p C\p m}\right) \,,\nn\\
\eea
where \eqref{result24} has been used in the last line.
Thus, we obtain
\bea\label{result34}
\cint \frac{z}{\sqrt{P_4}}dz &=& -\frac{m^2(1-q)^2-C(1+q)}{C}\frac{\p a_0}{\p C} \nn\\[3mm]
&&\quad - \frac{(m(1-q)^2-C(1+q))^2}{Cm(1-q)^2}\frac{\p^2a_0}{\p C\p m} + \frac{4Cq}{m(1-q)^2}\frac{\p^2a_0}{\p C\p m} \,.
\eea

The integral used in \eqref{ftt4} is computed as
\bea\label{result44}
\cint \frac{1}{z}\frac{dz}{\sqrt{P_4}} &=& \cint (Cz^2  + (m^2(1-q)^2-C(1+q))z + Cq)\frac{dz}{P_4^{3/2}} \nn\\[3mm]
&=& \cint (-2Cz^2 - (m^2(1-q)^2-C(1+q))z)\frac{dz}{P_4^{3/2}} \nn\\[3mm]
&=& \frac{4C}{m(1-q)^2}\frac{\p^2a_0}{\p C\p m} - \cint (m^2(1-q)^2-C(1+q))z\frac{dz}{P_4^{3/2}} \,,
\eea
where in the second equality the relation
\bea
0 &=& \cint \frac{P_4'}{P_4^{3/2}}dz \nn\\[3mm]
&=& \cint (3Cz^2 + 2(m^2(1-q)^2-C(1+q))z + Cq)\frac{dz}{P_4^{3/2}}
\eea
was used to replace the constant term.
\\

\section{Nekrasov Partition Function}\label{Nek}

We give a brief summary of the Nekrasov instanton partition function and present relevant results obtained from them. \cite{Nekrasov:2002qd}, \cite{Nekrasov:2003rj}

\subsection{$N_F=4$ theory}\label{af4}

The instanton part of the Nekrasov partition function for $\CN=2$ $U(2)$ gauge theory with four anti-fundamental hypermultiplets is of the form
\bea
Z_{\rm{inst}} = \sum_{\vec{Y}}q^{|\vec{Y}|}Z_{\rm{vec}}(\vec{a},\vec{Y})\prod_{i=1}^4Z_{\rm{afund}}(\vec{a},\vec{Y}, \mu_i) \,,
\eea
where $\vec{Y}=(Y_1,Y_2)$ is a pair of Young diagrams, $\vec{a}=(a_1,a_2)$ a pair of Coulomb branch parameters and $\mu_i$ denotes the mass of the hypermutiplet. The vector mutiplet and anti-fundamental hypermultiplet contributions are given by
\beq
Z_{\rm{vec}}(\vec{a},\vec{Y})= \prod_{i,j=1}^2\prod_{s\in Y_i}(a_{ij}-\e_1L_{Y_j}(s)+\e_2(A_{Y_i}(s)+1))^{-1}\prod_{t\in Y_j}(a_{ji}+\e_1(L_{Y_j}(t)+1)-\e_2A_{Y_i}(t))^{-1} \,,\nn
\eeq
\beq
Z_{\rm{afund}}(\vec{a},\vec{Y},\mu) = \prod_{i=1}^2 \prod_{s\in Y_i}(a_i+\e_1(m-1)+\e_2(n-1)+\mu) \,,
\eeq
where $a_{ij}\equiv a_i-a_j$, and the leg-length $L_{Y_i}(s)=\l_n'-m$ and the arm-length $A_{Y_i}(s) = \l_m-n$ are defined for a box $s$ at position $(m,n)$ in a Young diagram $Y_i = (\l_1\geq\l_2\geq\cdots)$, with its transpose $Y^t_i=(\l_1'\geq\l_2'\geq\cdots)$.

To compare with the results from matrix models, we consider $SU(2)$ gauge group with the Coulomb branch parameter $\vec{a}=(a,-a)$, noting that the contributions coming from the $U(1)$ factors are irrelevant for genus one correction. We define the free energy as
\bea
\CF &\equiv& -\e_1\e_2\log Z_{\rm{inst}}  \nn\\[2mm]
&=& \sum_{k,l=0}g_s^{2k}\e^l \CF_{k,l} \,.
\eea
For the case of $\mu_i=m$, which we considered in the matrix model computation, we calculate the free energy up to $g_s^2$ order.
\bea
 \CF_{0,0}&=& \frac{a^4+6 a^2 m^2+m^4}{2 a^2}q + \frac{13 a^8+100 a^6 m^2+22 a^4 m^4-12 a^2 m^6+5 m^8}{64 a^6}q^2
\nonumber\\[3mm]
&& + \frac{23 a^{12}+204 a^{10} m^2+51 a^8 m^4-48 a^6 m^6+45 a^4 m^8-28 a^2 m^{10}+9 m^{12}}{192 a^{10}}q^3
\nonumber\\[3mm]
&& + \frac{1}{32768 a^{14}}\left(2701 a^{16}+26440 a^{14} m^2+7164 a^{12} m^4-9000 a^{10} m^6\right.
\nonumber\\[3mm]
&&\qquad \left.+12190 a^8 m^8-13384 a^6 m^{10}+10908 a^4 m^{12}-5720 a^2 m^{14}+1469 m^{16}\right)q^4
\nonumber\\[3mm]
&& + \mathcal{O}(q^5)\,,\label{af004}\\[6mm]
\CF_{0,1} &=& -\frac{m \left(a^2+m^2\right)}{a^2}q -\frac{9 a^6m+11 a^4 m^3-9 a^2 m^5+5 m^7}{16a^6}q^2
\nonumber\\[3mm]
&&-\frac{38 a^{10}m+51 a^8 m^3-72 a^6 m^5+90 a^4 m^7-70 a^2 m^9+27 m^{11}}{96a^{10}}q^3
\nonumber\\[3mm]
&&-\frac{1}{4096 a^{14}} \left(1257 a^{14}m+1791 a^{12} m^3-3375 a^{10} m^5+6095 a^8 m^7-8365 a^6 m^9\right.
\nonumber\\[3mm]
&&\qquad\qquad\left. +8181 a^4 m^{11}-5005 a^2 m^{13}+1469 m^{15}\right)q^4\;+\;\mathcal{O}(q^5) \,,\label{af014}
\eea
\bea
\CF_{0,2} &=& \frac{a^4 + 6 a^2 m^2 + m^4}{8 a^4}q + \frac{9 a^8 + 64 a^6 m^2 - 70 a^4 m^4 + 40 a^2 m^6 + 21 m^8}{128 a^8}q^2 \nn\\[3mm]
&& + \frac{19 a^{12} + 147 a^10 m^2 - 300 a^8 m^4 + 470 a^6 m^6 -
   357 a^4 m^8 + 39 a^2 m^{10} + 110 m^{12}}{384 a^{12}}q^3 \nn\\[3mm]
&& + \frac{1}{32768 a^{16}}(1257 a^{16} + 10276 a^{14} m^2 - 28776 a^{12} m^4 +
   67460 a^{10} m^6 - 105630 a^8 m^8 \nn\\[3mm]
&&\qquad + 103308 a^6 m^{10} -
   43120 a^4 m^{12} - 15028 a^2 m^{14} + 18445 m^{16})q^4 \;+\; \CO(q^5) \,,\label{af024} \\[7mm]
\CF_{1,0} &=& \frac{m^2 (-a^2 + m^2)^3}{32 a^8}q^2 + \frac{m^2 (-a^2 + m^2)^3 (3 a^4 - 3 a^2 m^2 + 8 m^4)}{96 a^{12}}q^3 \nn\\[3mm]
&& + \frac{m^2 (-a^2 + m^2)^3 (235 a^8 - 432 a^6 m^2 + 1486 a^4 m^4 -
   1656 a^2 m^6 + 1647 m^8)}{8192 a^{16}}q^4 \;+\; \CO(q^5) \,. \nn\\
\eea
\\

\subsection{$N_F=3$ theory}\label{af3}

The instanton partition function for this case is
\beq
Z_{\rm{inst}} = \sum_{\vec{Y}}\L_3^{|\vec{Y}|}Z_{\rm{vec}}(\vec{a},\vec{Y})\prod_{i=1}^3Z_{\rm{afund}}(\vec{a}, \vec{Y}, \mu_i) \,,
\eeq
where $\L_3$ is the dynamical scale of the gauge theory. As in section \ref{f3}, we set $\mu_1=\mu_2=0$ and $\mu_3=m$. The free energy is computed as
\bea
 \CF_{0,0} &=& \frac{m}{2}\Lambda + \frac{a^2+m^2}{64 a^2}\Lambda^2 + \frac{a^4-6 a^2 m^2+5 m^4}{32768 a^6}\Lambda^4 + \cdots\,,
\\[7mm]
 \CF_{0,1} &=&
-\frac{\Lambda}{4} -\frac{m}{64 a^2}\Lambda^2 +\frac{m \left(3 a^2-5 m^2\right)}{16384\, a^6}\Lambda^4 + \cdots\,,\\[7mm]
\CF_{0,2} &=& \frac{m}{8 a^2}\L + \frac{a^2 - m^2}{128 a^4}\L^2 + \frac{m (-3 a^2 + 5 m^2)}{768 a^6}\L^3 \nn\\[3mm]
&& - \frac{2 a^4 - 25 a^2 m^2 + 21 m^4}{32768 a^8}\L^4 + \cdots\,,\\[7mm]
\CF_{1,0} &=& -\frac{a^2 - 2 m^2}{128 a^4}\L^2 + \frac{3 a^4 - 41 a^2 m^2 + 46 m^4}{32768 a^8}\L^4 + \cdots \,.
\eea
\\

\subsection{$N_F=2$ theory}\label{af2}

In this case, the instanton partition function is of the form
\beq
Z_{\rm{inst}} = \sum_{\vec{Y}}\L_2^{2|\vec{Y}|}Z_{\rm{vec}}(\vec{a},\vec{Y})\prod_{i=1,3}Z_{\rm{afund}}(\vec{a}, \vec{Y}, \mu_i) \,.
\eeq
By setting $\mu_1=\mu_3=m$ as in section \ref{f2}, we find the free energy as
\bea
\CF_{0,0} &=& \frac{a^2+m^2}{2a^2}\Lambda^2 + \frac{a^4-6 a^2 m^2+5 m^4}{64 a^6}\Lambda^4+\frac{5 a^4 m^2-14 a^2 m^4+9 m^6}{192\, a^{10}}\Lambda^6
\nn\\[3mm]
&& +\frac{5 a^8-252 a^6 m^2+1638 a^4 m^4-2860 a^2 m^6+1469 m^8}{32768\, a^{14}}\Lambda^8 + \cdots \,,
\eea
\bea
 \CF_{0,1} &=& -\frac{m}{2a^2}\Lambda^2 +\frac{m \left(3 a^2-5 m^2\right) }{32 a^6}\Lambda^4-\frac{m \left(5 a^4-28 a^2 m^2+27 m^4\right)}{192\,a^{10}}\Lambda^6
\nonumber\\[3mm]
&& +\frac{m \left(63 a^6-819 a^4 m^2+2145 a^2 m^4-1469 m^6\right)}{8192\, a^{14}}\Lambda^8 + \cdots \,,\\[7mm]
\CF_{0,2} &=& \frac{a^2 + m^2}{8 a^4}\L^2 - \frac{a^4 - 21 m^4}{128 a^8}\L^4 + \frac{5 a^6 - 14 a^4 m^2 - 123 a^2 m^4 + 220 m^6}{768 a^{12}}\L^6 \nn\\[3mm]
&& + \frac{-21 a^8 + 738 a^6 m^2 + 3080 a^4 m^4 - 19266 a^2 m^6 +
   18445 m^8}{32768 a^{16}}\L^8 + \cdots \,,\label{af022}\\[7mm]
\CF_{1,0} &=& \frac{a^4 - 3 a^2 m^2 + 2 m^4}{64 a^8}\L^4 + \frac{11 a^4 m^2 - 27 a^2 m^4 + 16 m^6}{192 a^{12}}\L^6 \nn\\[3mm]
&& + \frac{23 a^8 - 615 a^6 m^2 + 3895 a^4 m^4 - 6597 a^2 m^6 +
   3294 m^8}{16384 a^{16}}\L^8 + \cdots\,.
\eea

\end{document}